\documentclass[11pt]{article}

\usepackage{bbm}
\begin{document}

\renewcommand{\thefootnote}{\fnsymbol{footnote}}

\begin{titlepage}
\begin{center}
\hfill LTH 739 \\
\hfill {\tt hep-th/0703035}\\
\vskip 10mm

{\Large
{\bf 
Supersymmetric Black Holes in String Theory\footnote{This
article is based on an overview talk given at the 2nd Workshop
and Midterm Meeting of the RTN project Constituents, Fundamental
Forces and Symmetries of the Universe in Naples, October 9-13,
2006.}
}
}

\vskip 10mm

\textbf{Thomas Mohaupt}

\vskip 4mm

Theoretical Physics Division\\
Department of Mathematical Sciences\\
University of Liverpool\\
Liverpool L69 3BX, UK \\
{\tt Thomas.Mohaupt@liv.ac.uk}
\end{center}
 
\vskip .2in

\begin{center} {\bf ABSTRACT} \end{center}
\begin{quotation} \noindent
We review recent developments concerning supersymmetric
black holes in string theory. After a general introduction
to the laws of black hole
mechanics and to black hole entropy in string theory, we
discuss black hole solutions in $N=2$ supergravity,
special geometry, the black hole attractor equations and
the underlying variational principle. Special attention
is payed to the crucial role of higher derivative corrections.
Finally we discuss black hole partition functions and their
relation to the topological string, mainly from the supergravity
perspective. We try to summarize the state of art and 
discuss various open questions and problems.

\end{quotation}

\vfill

\end{titlepage}

\eject

\renewcommand{\thefootnote}{\arabic{footnote}}

\section{Introduction to black holes}

\subsection{The laws of black hole mechanics}

The laws of black hole mechanics \cite{BCH:73} imply that
black hole solutions of classical gravity are subject to relations
which are formally equivalent to the laws of 
thermodynamics. The zeroth law states that the so-called surface
gravity $\kappa_S$ of a stationary black hole
is constant over the event horizon,
\begin{equation}
\kappa_S = \mbox{const.}
\end{equation}
The surface gravity of a stationary black hole is the acceleration
of a static observer at the horizon, as measured by an observer
at infinity.\footnote{We refer the reader to \cite{Wald,FroNov} for an
extensive discussion of this important quantity. Both books cover 
all expects of black holes which are relevant in this section.}
The first law,
\begin{equation}
\delta M = \frac{\kappa_s}{8 \pi} \delta A + \omega \delta J 
+ \phi \delta q \;,
\end{equation}
relates a change of the mass $M$ of a stationary black hole to changes
of the area $A$, the angular momentum $J$ and the electric charge $q$. 
The parameters
multiplying the variations are the surface gravity $\kappa_S$, 
the angular velocity at the horizon $\omega$, and the electrostatic
potential at the horizon $\phi$. The second law,
\begin{equation}
\delta A \geq 0 \;,
\end{equation}
states that the total area of event horizons 
is non-decreasing. Finally, the third law states that the extremal 
limit $\kappa_S = 0$ cannot be reached in finite time by any
physical process. 

The laws suggest to identify surface gravity
with temperature and area with entropy, 
\begin{equation}
\kappa_S \sim T \;,\;\;\;A \sim S \;.
\end{equation}

In classical physics it does not make sense to assign a finite temperature
to a black hole, because it cannot emit radiation. 
However, once quantum effects
are taken into account a black hole of surface gravity $\kappa_S$ is
found to have the (Hawking) temperature \cite{Haw}\footnote{We use units
where $c=\hbar=G_N=1$.}
\begin{equation}
T = \frac{\kappa_S}{2 \pi} \;,
\label{HawkingTemperature}
\end{equation}
which fixes 
\begin{equation}
S = \frac{A}{4} \;.
\end{equation}
We will come back to this later, and for the time being we focus on the 
assumptions underlying the derivation of the laws of black hole
mechanics. The laws
are statements about {\em solutions} of the field equations, and in the
original proofs the Einstein equations are used explicitly. The
behaviour of matter is controlled by imposing a
suitable energy condition.\footnote{such as the dominant or the weak
energy condition, see \cite{Wald,FroNov} and original papers for details.}
While the zeroth and first law refer to stationary black holes,
the second law allows processes where black holes collide and
merge. To have control over the time evolution one needs to 
impose (in addition to the field equations and an energy condition),
that the space-time is `asymptotically predictable'.

It turns out that the zeroth and first law do not depend on 
details of the field equations. They can be proved for any
reparametrization invariant action, provided one makes the following 
assumptions: (i) the
field equations admit stationary black hole solutions,
which are either static, or axisymmetric and $t-\phi$ reflection
symmetric, (ii) the event horizon is a Killing horizon, and,
(iii) the space-time is globally hyperbolic \cite{Wald:93,RacWal:95}.
Let us explain these conditions in more detail. In the following 
the only restriction imposed on the action is that it is 
reparametrization invariant. In particular, the action might
contain higher derivative and especially higher curvature terms,
which are expected to be present in quantum-effective actions. 
Concerning condition (i): evidently, if the
action is restricted by nothing but general covariance, it is not
clear that it admits black hole solutions, which therefore needs to
be postulated. Moreover, the proofs of the zeroth and first law 
rely on symmetry properties, which in Einstein gravity
follow from the field equations, but need to be postulated when working 
in a
more general setting. Recall that a space-time is called stationary 
when it admits a time-like Killing vector field, which in adapted
coordinates takes the form $\xi = \partial_t$. Here $t$ is a timelike
coordinate along the integral lines of the Killing vector field.
A space-time is called static
if in addition the Killing vector field is hypersurface 
orthogonal. The latter statement is equivalent to the existence
of a coordinate transformation which removes
the mixed components $g_{ti}$ of the metric. 
For non-static stationary black holes
one needs to require that they are axisymmetric. The associated
Killing vector field is denoted $\partial_\phi$. Moreover it must
be symmetric under
simultanous reflection of the time coordinate $t$ and the angular
coordinate $\phi$.
This is called $t-\phi$ reflection symmetry
and is a well-known property of the Kerr solution. Concerning 
condition (ii): a Killing horizon is a hypersurface in space-time
where a Killing vector fields becomes null. In Einstein gravity all
event horizons of stationary black holes are Killing horizons.
It is not known whether this is true for general gravitational
actions, but the proofs of the zeroth and first theorem 
make use of this property. For static
black holes the `horizontal' Killing vector field is just the
one associated with time-translation invariance, $\xi = \partial_t$,
whereas
it takes the form $\xi = \partial_t + \omega \partial_\phi$ in the
non-static case. A surface gravity can be naturally defined for 
any Killing horizon, and its properties are essentially determined 
by the horizontal Killing vector field. Concerning condition (iii):
this means that the space-time has a Cauchy hypersurface and, hence,
a well defined initial value problem. With these assumptions
the zeroth law can be proved, without using the explicit form of the
field equations. The proof of the first theorem needs an additional
ingredient: the definition of the black hole entropy needs to
be generalized. The proof of the first law for general
actions\footnote{Here we refer to gravitational actions which are
reparametrization invariant and admit stationary black hole 
solutions, but are not restricted otherwise.}
relies on the observation that 
all the quantities which are varied, i.e., mass,
entropy, angular momentum and charge can be expressed as surface
charges. This is well known for the electric charge, which 
can be measured by the flux of the electric field through a two-sphere
at spatial infinity. But it is also true for entropy, mass and 
angular momentum. The reason is that the Noether current associated
with reparametrisations is itself a total derivative, i.e., it
can be written as the divergence of an antisymmetric Noether potential,
$J^\mu = \nabla_\nu Q^{\mu \nu}$ \cite{Wald:90}. 
Therefore the associated Noether
charges, which a priori involve integration over a spatial 
hypersurface, can be re-written as surface charges. The basic observation
underlying the 
proof of the first law is that the Noether charge associated with
the horizontal Killing vector field $\xi = \partial_t + \omega \partial_\phi$
does not change when deforming a given stationary solution infinitesimally
into another stationary solution \cite{Wald:93}. 
The Noether charge receives
two contributions, one from infinity, the other from the horizon. Since the
total charge is conserved, changes in the two contributions must balance
against each other. The contribution from infinity contains two terms:
one proportional to $\partial_t$, which gives the change in mass, another
one proportional to $\partial_\phi$, which gives the change in angular
momentum. Therefore the contribution from the horizon is interpreted
as the change in entropy. This can be used to define the black hole
entropy for general gravitational actions such that the 
first law of black hole mechanics holds:
\begin{equation}
\delta M = \frac{\kappa_S}{2 \pi} \delta S + \omega \delta J +
\phi \delta q \;,
\end{equation}
where the entropy is given by Wald's formula \cite{Wald:93}
\begin{equation}
S =  2\pi \int_H \tilde{Q}^{\mu \nu} d \Sigma_{\mu \nu} \;.
\label{Wald}
\end{equation}
The tensor field $\tilde{Q}^{\mu \nu}$ is related to the Noether 
potential $Q^{\mu \nu}$ by scaling out the surface gravity
$\kappa_S$. The integral can be taken over any spatial
section $H$ of the event horizon \cite{Myers}. 
The Noether charge is linear in the Killing vector field
and its derivatives. It has been shown that it can be expressed 
in terms of the variational derivative of the Lagrangian 
with respect to the Riemann tensor by \cite{WaldVar}
\begin{eqnarray}
S &=& 2 \pi \int_H \left(
\frac{\partial L}{\partial R_{\mu \nu \rho \sigma}} - 
\nabla_\alpha \frac{\partial L}{\partial 
(\nabla_\alpha R_{\mu \nu \rho \sigma})} +
\nabla_{(\alpha} \nabla_{\beta)} \frac{\partial L}{\partial 
(\nabla_{(\alpha} \nabla_{\beta)}  R_{\mu \nu \rho \sigma})}
 \right. \nonumber \\
& &\left. - \nabla_{(\alpha} \nabla_{\beta} \nabla_{\gamma)}
\frac{ \partial L}{ \partial (  \nabla_{(\alpha} 
\nabla_{\beta} \nabla_{\gamma)}
R_{\mu \nu \rho \sigma}) } + \cdots 
\right) \varepsilon^{\mu \nu} \varepsilon^{\rho \sigma} \, 
\sqrt{h} \, d^2 \Omega 
\;.
\end{eqnarray}
Here $\varepsilon^{\mu \nu}$ is the normal bivector of $H$, and 
$h$ denotes the induced metric. It is easy to check that this
formula reproduces the area law $S = \frac{A}{4}$ if the only
term which contributes is the Einstein-Hilbert term. However, 
if the action contains further curvature terms, there are
modifications in general.

At this point it is natural to wonder whether the second law can
be proved for general gravitational actions. Though it has been
observed  in the literature that the 
Noether charge is well defined and natural for non-stationary
space-times \cite{WaldVar}, there is no general proof of the second law in this
setting. Since one has to admit dynamical processes such as black hole
collisions and black hole fusion, one might also question whether 
the second law should hold for any action, or whether it 
singles out a subclass of `physically reasonable' ones. 
We will comment on the second and third law for quantum black holes
in the next section.

\subsection{Quantum mechanical aspects of black holes}

So far our discussion appeared to be purely classical, in that we 
studied the properties of stationary black hole solutions associated
to some Lagrangian. But as is obvious from the very name, black holes
can only absorbe matter and radiation, and therefore it does not
make sense to assign to them a finite temperature. This changes once
quantum effects in the matter sector are taken into account. This
can be done using quantum field theory in curved space-time, where
gravity is taken as a classical background field, while matter is
described by quantum field theory \cite{Wald2,BirDav,FroNov}. 
The Hawking effect \cite{Haw} can be 
derived by either taking space-time to describe the collapse of a
mass distribution, or by taking a stationary black hole background.
In both cases one finds outgoing radiation which allows to assign
the (Hawking) temperature (\ref{HawkingTemperature}) to the black hole.
Combining this with the first law one concludes that the black hole
should have the entropy
\begin{equation}
S = \frac{A}{4} + \mbox{corrections from further curvature terms} \;,
\end{equation}
where we took already into account that the area law is modified when
further curvature terms are present in the Lagrangian.\footnote{
Since space-time plays the role of a classical background field, 
the derivation of the Hawking effect is independent of the details of
the gravitational field equations.} 

Of course, quantum field theory in curved space-time is only an
approximation of a full theory of quantum gravity.\footnote{Ultimately,
this approximantion is inconsistent, see for example \cite{Kiefer}
for a discussion.} 
The full quantum
treatment of black holes poses a lot of challenges, both conceptually
and technically. The success of any proposed theory of quantum gravity
can be judged by its capability to provide answers. Here we will focus
on one particular aspect, the so-called entropy problem, and we will
review what string theory has to say about this. We will not be able
to give a complete overview, but focus on one line of thought, which
can be tested with high precision.

The entropy $S$ appearing in the laws of black hole mechanics 
should be 
interpreted as a thermodynamical or macroscopic entropy, denoted
$S_{\rm macro}$. More generally,
we think about the description of black holes provided by a
general gravitational action as a coarse-grained, macroscopic, effective
description of an underlying microscopic quantum theory. The microcsopic
or statistical entropy $S_{\rm micro}$ is then given by 
\begin{equation}
S_{\rm micro} = \log d(M,J,q) \;,
\end{equation}
where $d(M,J,q)$ is the number of microstates of a black hole
with the macroscopic parameters $M,J,q$. We expect that 
macroscopic and microscopic entropy agree, at least in the
thermodynamical limit, which we identify with the semiclassical limit
where the macroscopic parameters (if non-vanishing\footnote{We 
admit black holes which do not carry angular momentum or 
electric charge.}) become large.  

The reason why we reviewed the definition of the (macroscopic) entropy
for generally covariant theories is that we want to make
a detailed comparison between both entropies, including subleading
corrections. The Lagrangians which we use in the macroscopic description
are taken to be effective Lagrangians derived from an underlying
microscopic theory by integrating out the massive modes and 
expanding in derivatives. Such effective Lagrangians are expected to 
be covariant, and to contain higher derivative and in particular
higher curvature terms.  
Any quantum theory of gravity should provide a way for 
calculating effective Lagrangians. In string theory effective
Lagrangians are computable in principle (UV finiteness),
and they have been computed in practice in several cases. 

On the microscopic side one needs to identify and count the
microstates which give rise to the same coarse-grained macrostate. The
basic idea will be explained in the next section. Before turning to 
this, let us also point out that there are further goals beyond just
matching the macroscopic and microscopic entropy. Ultimately one would
like to derive the macroscopic properties of black holes from the
underlying microscopic theory. One step in this direction is the 
OSV conjecture \cite{OSV}, which defines a black hole partition function 
and relates it to the partition function of the topological string. 
We will come back to this at the end of this article. 

Finally, let us comment on the second and third law in the quantum
realm. Since Hawking radiation extracts energy from a black hole,
it will shrink, thus violating the second law of black hole
mechanics. This complements
the observation that the presence of black holes leads to 
a violation of the classical second law of thermodynamics, because
entropy could be reduced by moving matter adiabatically into 
black holes. This is an independent motivation for assigning 
an entropy to black holes \cite{Bekenstein}. The generalized 
second law of thermodynamics, which is expected to hold quantum
mechanically, states that 
the total entropy obtained by adding thermodynamical and black hole
entropy is non-decreasing. 

Like the third law of thermodynamics, the third law of black hole
mechanics has different versions. The strong version states that
in the extremal limit (zero temperature) the entropy vanishes. 
This version is already violated in thermodynamics \cite{Kiefer}, 
and for black holes
it is at odds with the fact that extremal black holes have a finite
horizon. While the application of Euclidean methods to extremal
black holes yields a vanishing entropy \cite{ZeroEntropy}, it has been pointed 
out in \cite{LouKie} that the extremal limit does not commute
with quantization. If the extremal limit is taken within the
quantum theory, the entropy is finite. This is consistent with
results from string theory where extremal black holes have a 
finite microscopic entropy \cite{StrVaf}. The weak version of the third
theorem states the extremal limit cannot be reached in finite time
by any physical process. We refer to \cite{Wald,Kiefer} for a
more detailed discussion. For our purposes the third law will not
be of great importance. We will be mainly concerned with extremal
black holes, but these are used as theoretical laboratories 
to test ideas, not as a realistic description of astrophysical
black holes.\footnote{Charged black holes are unstable with 
respect to charge superradiance, and therefore black holes
are expected to discharge quickly and to settle down to a 
Kerr-like or Schwarzschild-like stationary state \cite{FroNov}.}
Therefore it not important whether the extremal limit can 
actually be reached in finite time.

\subsection{The string-black hole correspondence}

The string-black hole correspondence introduced in \cite{Sus:93}
is a heuristic but very general way to understand the entropy 
of black holes. The basic idea is that black hole microstates 
simply are states of highly excited massive strings. Consider,
for concreteness, strings moving in a flat four-dimensional 
space-time (neglecting that quantum consistency requires additional
dimensions).  Since string theory contains gravity, one might
wonder why it is allowed to neglect the backreaction
of the strings on space-time. The answer is that at asymptotically
small string coupling the string scale $\sqrt{\alpha'}$ is much
larger than the Schwarzschild radius of the string, $r_S$. Since 
the Schwarzschild radius sets the scale for the backreaction on 
space-time, it is consistent to treat space-time as a flat 
background in the string perturbative regime $\sqrt{\alpha'} \gg 
r_S$. However, if we sufficiently increase the mass (at fixed coupling)
or the coupling (at fixed 
mass), then $r_S$ will grow relative
to $\sqrt{\alpha'}$ and, ultimately it will become comparable. While 
not much is known about what happens precisely in this regime, 
we can go to the opposite limit where $\sqrt{\alpha'} \ll r_S$.
Now the Schwarzschild radius is much larger than the string scale,
and stringy effects are suppressed. Therefore we can 
apply classical gravity and conclude that the string state should
be described as a black hole, because it sits within its Schwarzschild
radius. Note that $\sqrt{\alpha'} \ll r_S$ means that higher 
derivative terms in the effective Lagrangian are suppressed. We call this
regime the semi-classical gravity regime. 

Assuming that the interpolation is allowed, we can count the number
of states of a string of given mass and compare it to the entropy
of a black hole of the same mass. We need a few formulae. First 
note that in four dimensions the Regge slope $\alpha'$ and 
Newton's constant $G_N$ are related by
\begin{equation}
G_N = g_S^2 \alpha' \;,
\end{equation}
where $g_S$ is the four-dimensional string coupling. For large 
excitation numbers $n\in \mathbbm{N}$, the mass $M$ of a string state
goes like
\begin{equation}
\alpha' M^2 \simeq n \;.
\end{equation}
The number of states at given $n$ grows like $\exp(\sqrt{n})$ \cite{GSW},
so that the microscopic (statistical) entropy is (to leading order) 
\begin{equation}
S_{\rm micro} = \log d(n) \simeq \sqrt{n} \;.
\label{Sstatstring}
\end{equation}
Using the relation between $G_N$ and $\alpha'$, the Schwarzschild 
radius of a string state of level $n$ is 
\begin{equation}
r_S \simeq g_S^2 \sqrt{n} \sqrt{\alpha'} \;.
\end{equation}
Now we can read off that the string perturbative regime corresponds
to $g_S^2 \sqrt{n} \ll 1$, while the semiclassical gravity regime
corresponds to $g_S^2 \sqrt{n} \gg 1$.\footnote{Since the effective
action which we use in the semi-classical gravity regime is computed
using string perturbation theory we need that the coupling is small, 
$g_S \ll 1$. This is not a problem, since we are interested in states
with large $n$.}
We can also compare the statistical entropy  (\ref{Sstatstring})
of string states to the Bekenstein-Hawking entropy $S_{\rm macro}\simeq
A \simeq r_S^2$ of a black hole of the same mass.
While both entropies disagree in general, we observe that they 
are equal within one order of magnitude when the Schwarzschild radius
equals the string scale:
\begin{equation}
g_S^2 \sqrt{n} \simeq 1 \Leftrightarrow r_S \simeq \sqrt{\alpha'} 
\Rightarrow S_{\rm micro} \simeq S_{\rm macro} \;.
\label{Matching1}
\end{equation}
This is of course the regime where we cannot describe the black hole
in detail. But the matching of the entropies suggests
that there is a transition between the two descriptions in this
regime. This might be a phase transition or a smooth crossover,
and is possibly related to the Hagedorn transition.\footnote{See 
\cite{HagedornTransition}
for recent work on the Hagedorn transition and black holes.}  
This scenario makes a prediction for the
final state of black hole evaporation: a black hole looses mass
through Hawking radiation until its size reaches the string scale
where it converts into a highly excited string state. Since there
is precisely the right number of string states to account for the 
black hole entropy, one expects that the time evolution
is unitary and that there is no information loss.

While this picture of black holes is very general, it is
heuristic and begs many questions.
A priori it is not clear whether the number 
of states should be conserved when we interpolate between 
the two regimes. Also, the matching (\ref{Matching1}) holds 
only up to factors of order unity, and there are subleading corrections
on both sides. One way to improve the situation is to consider special 
supersymmetric (BPS) states, where both the macroscopic entropy 
$S_{\rm macro}$ and the microscopic entropy 
$S_{\rm micro}$ can be computed to high precision
in their respective regimes. The interpolation between both regimes
is also more reliable in this case, because BPS states saturate
the mass bound implied by the supersymmetry algebra (more later). 
While the original string-black hole correspondences only invokes
fundamental string states, there are other candidates for black hole
microstates in string theory. One important class are D-branes
whose excitations are described by the open
strings ending on it. The first successful quantitative 
derivation of black hole entropy was based on D-branes \cite{StrVaf}. 
A unified formulation of the string-black hole
correspondence which encompases both fundamental string states and
D-branes was formulated in \cite{HorPol}.

We should also stress that the space-time into which the fundamental 
strings or D-branes are embedded need not to be flat. It can be
any curved (on-shell) string background. If we want to 
describe four-dimensional black holes
we should consider space-times where the extra dimensions have been
compactified. While this enlarges the number of 
possible microscopic descriptions of black holes, one feature 
that we saw in the above example remains: one needs to interpolate
between two regimes. The state counting is done in a regime where
the feedback of strings and D-branes on the background 
can be neglected, while the description as a four-dimensional 
black hole with an event horizon is valid when we can use a four-dimensional
effective action. The interpolation 
between these two regimes involves the variation of 
the string coupling and possibly of geometrical moduli.

\section{Black holes in $N=2$ supergravity}

\subsection{BPS states and BPS solitons}

Let us review some standard facts about supersymmetry
algebras and their representations \cite{WessBagger,West}.
The $N$-extended four-dimensional supersymmetry algebra takes the
following form:
\begin{equation}
\{ Q_\alpha^A, Q^{+B}_\beta \} = 2 \sigma^\mu_{\alpha \beta} 
\delta^{AB}  P_\mu\;,\;\;\;
\{ Q_\alpha^A, Q^{B}_\beta \} = \epsilon_{\alpha \beta} Z^{AB}  \;,
\end{equation}
where $A,B, \ldots = 1, \ldots, N$ labels the supercharges, which
we have taken to be Majorana spinors. $Z^{AB}$ is a complex
antisymmetric matrix of central operators. Using R-symmetry
rotations it can be skew-diagonalized. The eigenvalues, called
central charges are constant on irreducible representations by
Schur's lemma. The same is true for the mass $M^2 = - P_\mu P^\mu$.
Using the algebra one can show that the mass is bounded by
the central charges:
\begin{equation}
 M^2 \geq |Z_1|^2 \geq |Z_2|^2 \geq \cdots \geq 0 \;.
\end{equation}
If one or several of the inequalities for $M$ are saturated,
then part of the supercharges operate trivially. As a result
the corresponding representation is smaller than a generic
massive representation. Such representations are called
BPS representations, also shortened representations or
supersymmetric representations (since the states are invariant
under part of the supersymmetry transformations). The most
extreme case are the short or $\frac{1}{2}$-BPS representations, 
where all central charges are equal:
$M^2 = |Z_1|^2 = |Z_2|^2 = \cdots$ Such representations are 
invariant under half of the supersymmetry transformations
and have as many states as massless representations. 

BPS states cannot only be realized as one-particle representations
on the Hilbert space, but also as finite energy solutions of the
field equations. Such objects are called BPS solitons. By soliton
we refer to stationary, finite energy solutions $\Phi_0$ to the equations
of motion,\footnote{Here $\Phi$ is a collective notation for all
fundamental fields appearing in the Lagrangian, and $\Phi_0$ is
a particular field configuration which solves the equations of motion.} 
which are regular in the sense of not having naked 
singularities (admitting black holes). Note that due to the finite
energy condition a soliton must approach the vacuum asymptotically.
Since its energy density is effectively localized in space, it
is considered as a particle-like object. BPS solitons (also called
supersymmetric solitons) are in 
addition invariant under a subset of the supersymmetry transformations.
I.e., the supersymmetry transformation parameters $\varepsilon$ 
can be chosen such that the field configuration $\Phi_0$ is invariant:
\begin{equation}
\delta_{\varepsilon} \left. \Phi \right|_{\Phi_0} = 0 \;.
\end{equation}
The corresponding transformation parameters $\varepsilon$ are called
Killing spinors. 

The extremal Reissner-Nordstrom black hole is a standard example
for a BPS soliton \cite{Gibbons}. Since it is the prototype for the supersymmetric
black holes to be considered later, let us review its key features.
Einstein-Maxwell theory is the bosonic part of $N=2$ supergravity.
Therefore the extremal Reissner-Nordstrom black hole is a solution
to the full field equations of $N=2$ supergravity. The two
gravitini, which provide the fermionic degrees of freedom
are identically zero in this solution. Since the $N=2$ supersymmetry
algebra has eight real supercharges (two Majorana spinors), there
can be at most eight linearly independent Killing spinors. The
extremal Reissner-Nordstrom black hole has four Killing spinors
and is therefore a $\frac{1}{2}$-BPS solution. At infinity the
solution approaches the Minkowski vacuum, which is a supersymmetric
ground state and therefore invariant under all supersymmetry transformations.
The eight linearly independent Killing spinors can be taken to 
be constant.\footnote{Note that we have to distinguish between
genuine symmetries (rigid symmetries) and reparametrisations
(local symmetries). Like Killing vectors, Killing spinors correspond
to genuine symmetries. Their explicit form depends on the 
chosen coordinates. If we describe flat space in terms of 
Cartesian coordinates, the Killing spinors are constant.} 
A doubling of unbroken supersymmetries also occurs at the event
horizon. The asymptotic near-horizon solution is the so-called
Bertotti-Robinson solution. The geometry is $AdS^2 \times S^2$,
while the gauge field is covariantly constant. This solution
has eight Killing spinors, and can therefore be considered 
as an alternative supersymmetric vacuum. We observe a feature
which is typical for two-dimensional solitons: the solution
interpolates between two vacua. This is understandable,
since the extreme Reissner-Nordstrom
solution is spherically symmetric and effectively depends
only on one spatial coordinate, the radial one.

\subsection{Special geometry}

We will be interested in BPS black holes in four-dimensional
string compactifications with $N=2$ and $N=4$ supersymmetry.
The compactification of the heterotic string on $K3 \times T^2$
or of the type-II string on a Calabi-Yau threefold gives
$N=2$ supergravity plus some number $n$ of vector multiplets 
plus further matter multiplets which are irrelevant for
our purposes. The compactification of the heterotic string on
$T^6$ or of the type-II string on $K3 \times T^2$ gives
$N=4$ supergravity plus some number $n$ of vector multiplets.
We will describe the $N=4$ theory using the $N=2$ formalism.
Therefore we need to discuss $N=2$ vector multiplets. 

The main tool for handling $N=2$ vector multiplets is the
so-called special geometry discovered in \cite{deWVPr}. While
the Lagrangian is complicated, all couplings are encoded in
a single holomorphic function, the prepotential. This results
from an invariance of the field equations under $Sp(2n+2,\mathbbm{R})$
transformations, which generalize the electric-magnetic duality
rotations of Maxwell theory. Stringy symmetries, such as 
T-duality and S-duality, form a discrete subgroup of this
symplectic group. 

Let us provide some details. The $N=2$ gravity multiplet 
contains the vielbein (graviton), a doublet
of Majorana gravitini, and a vector
field called the graviphoton. A vector 
multiplet contains a vector field, a doublet of spinors and
a complex scalar. The bosonic Lagrangian is a generalized
Einstein-Maxwell Lagrangian plus a scalar sigma-model:
\newpage
\begin{eqnarray}
8 \pi e^{-1} {\cal L}_{\rm bos} &=& - \frac{R}{2} - 
g_{A\overline{B}} (z, \overline{z}) \partial_\mu z^A \partial^\mu 
\overline{z}^{\overline{B}} \nonumber \\ 
&& + \frac{i}{4} 
\overline{\cal N}_{IJ}(z, \overline{z})
F^{-I}_{\mu \nu} F^{-I| \mu \nu} 
- \frac{i}{4} {\cal N}_{IJ} (z, \overline{z}) 
F^{+I}_{\mu \nu} F^{+I| \mu \nu}  \;.
\end{eqnarray}
Here $z^A$, $A=1,\ldots,N$ are the scalar fields, and the 
field dependent coupling $g_{A\overline{B}}(z,\overline{z})$
can be interpreted as the metric of the scalar manifold
${\cal M}_{\rm VM}$. $F^{\pm I}_{\mu \nu}$ are the
(anti-)selfdual parts of the field strength of the
$n+1$ vector fields, $I=0,1,\ldots,n$. The gauge couplings
${\cal N}_{IJ}(z,\overline{z})$ are field-dependent. 

To make the $Sp(2n+2,\mathbbm{R})$ invariance of the field
equations manifest, one defines the dual field strength
by
\begin{equation}
G^{\pm}_{I|\mu \nu} \simeq \frac{\partial {\cal L}}{\partial F^{\pm I |\mu \nu}} \;.
\end{equation}
These are of course dependent fields, but they are useful 
because $(F^{\pm I}_{\mu \nu}, G^{\pm}_{I|\mu \nu})^T$ transforms
linearly under $Sp(2n+2,\mathbbm{R})$. Such quantities are called
symplectic vectors. Another symplectic vector is provided 
by the magnetic and electric charges, $(p^I, q_I)^T$. This is clear
because the charges are obtained by integrating the (dual)
field strength over an asymptotic two-sphere:
$p^I = \oint F^I_{\mu \nu} d \sigma^{\mu \nu}$, 
$q_I = \oint G_{I|\mu \nu} d\sigma^{\mu \nu}$. 
The $Sp(2n+2,\mathbbm{R})$ 
transformations
are not only invariances of the gauge field equations but of
the full field equations. To describe the scalars in a covariant
way, one does not work with the physical
scalar fields $z^A$, but with a symplectic vector $(X^I, F_I)^T$.
Here $X^I$ are homogenous coordinates on the scalar manifold
${\cal M}_{\rm VM}$, i.e. $z^A = \frac{X^A}{X^0}$. The prepotential $F$,
which encodes all vector multiplet coulings, is a function of the $X^I$.
Local $N=2$ supersymmetry implies that it is holomorphic and 
homogenous of degree two:
\begin{equation}
F(\lambda X^I) = \lambda^2 F(X^I) \;,\;\;\;\lambda \in \mathbbm{C}^* \;.
\end{equation}
The quantities $F_I$ which complete the symplectic vector 
$(X^I, F_I)^T$, are the components of the gradient of $F(X^I)$:
\begin{equation}
F_I = \frac{\partial F}{\partial X^I} \;.
\end{equation}

The structure of the scalar sector becomes transparent when 
constructing the theory using the superconformal calculus.
Here one starts with $n+1$ vector multiplets with scalars $X^I$
and imposes that
the theory is superconformally invariant. While (rigid) 
$N=2$ supersymmetry implies the existence of a holomorphic
prepotential, superconformal invariance imposes in addition
that $F$ must be homogenous of degree two. The next step
is to make the superconformal symmetry local. The necessary
covariantization introduces gauge fields which belong to  the
so-called Weyl multiplet.\footnote{This description is somewhat
simplified. For more details see \cite{Habil} and
references therein.} The resulting
theory turns out to be gauge equivalent to $N=2$ Poincar\'e
supergravity. This means that one can gauge fix the additional
symmetries to obtain a Poincar\'e supergravity
theory. Conversely, a Poincar\'e supergravity theory can be made
superconformal by simultanously introducing new symmetries
and new fields which act as compensators. The gauge fixing
imposes one (complex) relation on the superconformal scalars $X^I$,
which can then be expressed in terms of the $n$ independent
physical scalars $z^A$. In contrast, all $n+1$ gauge fields
remain independent, although one of them becomes part of
the Poincar\'e gravity multiplet, so that only $n$ 
vector multiplets remain. The other physical degrees
of freedom of the Poincar\'e gravity multiplet,
namely the graviton and two gravitini are provided by
the Weyl multiplet. 

The scalar geometry can be cast in the following form,
as explained in \cite{ACD}: while the physical scalars
$z^A$ are coordinates on a complex $n$-dimensional
K\"ahler manifold ${\cal M}_{\rm VM}$, the scalars $X^I$
are coordinates on a complex cone ${\cal N}_{\rm VM}$
over ${\cal M}_{\rm VM}$. This cone is the scalar manifold
of the rigid superconformal theory used as the starting point
in the superconformal calculus. The existence of a holomorphic
prepotential is equivalent to the local existence of a
holomorphic Lagrangian immersion $\phi = dF$ of the cone
into the flat complex symplectic vector space 
$T^* \mathbbm{C}^{n+1} \simeq \mathbbm{C}^{2n+2}$.
The quantities $(X^I, F_I)^T$ can be interpreted 
as symplectic coordinates on $T^* \mathbbm{C}^{n+1}$.
${\cal N}_{\rm VM}$ is generically immersed into this
space as a graph. Therefore one can take the $X^I$
as coordinates on ${\cal N}_{\rm VM}$, and the $F_I$
become functions of the $X^I$ along the immersed
space. Since the immersion is Lagrangian, the $F_I$
form the gradient of some function $F$, which generates
the immersion.\footnote{For special choices of coordinates,
the $X^I$ might be dependent along the immersed submanifold.
Then they do not provide coordinates on the submanifold,
and the $F_I$ cannot be written as the gradient of a
prepotential. However, one can always make a symplectic
change of coordinates on $T^* \mathbbm{C}^{n+1}$ such that the 
$X^I$ are coordinates on the immersed space
and then a prepotential exists. This phenomenon has been
discussed in \cite{CDFV,CRTV}.}
All geometric data of the cone, i.p. its K\"ahler metric and
its K\"ahler form can be obtaind by pulling back the
standard hermitean form of $T^* \mathbbm{C}^{n+1}$ 
with $\phi$. As a result all data can be expressed 
through the prepotential. In particular the K\"ahler
metric on ${\cal N}_{\rm VM}$ has a holomorphic
prepotential. 
Note that the relation between
${\cal N}_{\rm VM}$ and $T^*\mathbbm{C}^{n+1}$ applies
irrespective of whether the prepotential is homogenous or not.
Manifolds which can be obtained through this construction
as holomorphic, Lagrangian immersions into a complex symplectic
vector space
are called affine special K\"ahler manifolds\footnote{An 
intrinsic defintion of these manifolds was given in \cite{Freed}.}. 
They are the scalar manifolds of 
(not necesserily superconformal) rigid $N=2$ vector multiplets.

If one imposes superconformal invariance,
the prepotential must be  
homogenous of degree two. Geometrically, this implies that
the affine special K\"ahler manifold 
${\cal N}_{\rm VM}$ is a complex cone. One can then perform
a $\mathbbm{C}^*$-quotient and obtain a new K\"ahler 
manifold ${\cal M}_{\rm VM}$, which is the basis of the 
complex cone.  Such manifolds are called projective special
K\"ahler manifolds, and they are the scalar target spaces
of $N=2$ vector multiplets coupled to Poincar\'e
supergravity. The K\"ahler potential of the metric
on ${\cal M}_{\rm VM}$, in fact all couplings
in the Poincar\'e supergravity Lagrangian, can be
expressed in terms of the prepotential $F$. 

A natural realization of special geometry is provided
by the moduli space of Calabi-Yau threefolds $X$. In
string theory this occurs, for example, when 
compactifying type IIB string theory on $X$.
Then the scalar manifold of the 
physical vector multiplet scalars $z^A$ is the 
moduli space of complex structures of $X$,
while the cone parametrized by the $X^I$ is obtained
by combining deformations of the complex structure
with those of the holomorphic top form.\footnote{
Physical quantities only depend on the complex
structure, but not on the explicit choice of the
holomorphic top form. Thus the additional deformations
are gauge transformations.}  The
complex vector space into which the resulting
cone can be embedded is the complex middle
cohomology $H^3(X,\mathbbm{C})$. Its coordinates
$(X^I, F_I)$ are the periods of the holomorphic
top form.

\subsection{The black hole attractor mechanism}

We now turn to $\frac{1}{2}$-BPS solutions of 
$N=2$ supergravity. 
As we will see special geometry is extremely useful in
finding and analysing solutions. We only consider 
static, spherically symmetric solutions which describe
a single extremal black hole. Then the line element takes
the form
\begin{equation}
ds^2 = - e^{2g(r)} dt^2 + e^{2 f(r)} ( dr^2 + r^2 d \Omega^2)
\end{equation}
and supersymmetry requires in addition that 
$g(r) = -f(r)$ \cite{CdWKM:2000}. Moreover each gauge field
has only two independent components, corresponding to static, radially
symmetric electric and magnetic fields. The main new feature compared
to the extremal Reissner-Nordstrom black hole is that there are 
scalar fields, $z^A(r)$. To maintain symplectic covariance
one can work with the conformal scalars $X^I(r)$. In fact, it is 
convenient to rescale the $X^I(r)$ such that they become invariant
under the radial $U(1)$-part of the $\mathbbm{C}^*$-transformations. 
This can be done using the symplectic function 
\begin{equation}
Z = p^I F_I - q_I X^I \;,
\end{equation}
which transforms with the same phase as $X^I$. In an asymptotically
flat background this function agrees with the central charge of
the $N=2$ supersymmetry algebra, when evaluated at infinity. 
Therefore it is often simply called the central charge. Note
however that $Z$ is a function, which through the scalar fields
depends on space-time, i.e. $Z =Z(r)$ for the
backgrounds under consideration. The $U(1)$-invariant scalars are
then defined by $Y^I(r) = \overline{Z}(r) X^I(r)$ \cite{CdWKM:2000}. 
Note that 
the physical scalars are given by $z^A(r) = Y^A(r)/Y^0(r)$. 

The resulting solutions \cite{BLS,CdWKM:2000} have two interesting
asymptotic regimes. The first is $r\rightarrow \infty$, where
they become asymptotically flat. In this limit the scalars
approach arbitrary values $z^A \rightarrow z^A(\infty) \in 
{\cal M}_{\rm VM}$. Together with the $2n+2$ charges $p^I, q_I$
there are $4n+2$ real integration constants.
The second asymptotic regime is the event horizon, 
$r \rightarrow 0$. Here the scalars do not take arbitrary
values, but specific fixed point values which are determined
by the charges. This is the celebrated black hole attractor mechanism
\cite{FKS}. The reason for this behaviour is that if the solution
is to remain regular at the horizon, it has to approach the
Bertotti-Robinson solution $AdS^2 \times S^2$, which has eight
Killing spinors and is fully supersymmetric. The fact that the
scalars take special fixed point values follows from gradient flow
equations implied by the gaugino variations \cite{FKS}. It 
also follows by imposing full supersymmetry, irrespective
of whether the resulting Bertotti-Robinson solution is global, 
or the near horizon asymptotics of a black hole \cite{CdWKM:2000}.
Alternatively, the attractor mechanism can be derived from
the equations of motion, for example by studying the
motion of a test particle in the near horizon geometry \cite{FGK}.
This is governed by an effective `black hole potential', which
is extremized at the horizon. Equivalently, we might think
about the Bertotti-Robinson solution as a flux compactification
from four to two dimensions. Since the $S^2$ factor is not
Ricci flat, flux (a covariantly constant gauge field) has to be 
turned on to solve the Einstein equations. This induces a
potential for the scalars, which is extremized at the horizon.
The latter  arguments 
rely on the equations of motion and do not depend on supersymmetry.
They therefore suggest that the attractor mechanism can also be realized 
for non-supersymmetric black holes. This is indeed true, as we will
review below.

The values of the scalar fields at the horizon are determined
by the so-called attractor equations (or stabilization equations),
which in our conventions take the following form:
\begin{equation}
\left( \begin{array}{c}
Y^I - \overline{Y}^I \\
F_I - \overline{F}_I \\
\end{array}
\right)_* = i 
\left( \begin{array}{c}
p^I \\
q_I \\
\end{array}
\right) \;,
\label{AttractorEqsI}
\end{equation}
where $*$ denotes evaluation at the horizon.
Observe that this is an equation between two symplectic vectors.

As a consequence, the black hole entropy only depends on the
(discrete) charges, but not on the (continuous) moduli $z^A(\infty)$.
The area of the event horizon is proportional to 
$|Z|^2_*$, where the symplectic function 
$Z = p^I F_I (X) - q_I X^I$ is now evaluated on the event
horizon. 
As a result the macroscopic entropy is a symplectic function:
\begin{equation}
S_{\rm macro} = \frac{A}{4} = \pi |Z|^2_* = 
\pi | p^I F_I(X) - q_I X^I |^2  =
\pi  \left( p^I F_I (Y) - q_I Y^I \right)_*  \;.
\end{equation}

\subsection{The black hole variational principle \label{Sect:VarPri}}

Shortly after the discovery of the black hole attractor
mechanism, it was observed that the attractor equations
follow from a variational principle \cite{BCdWKLM}.
The importance of this observation was only appreciated
much later, after the work of \cite{OSV} uncovered a 
direct link between black hole entropy and the topological
string partition function.

The variational principle is based on the following 
`entropy function'\footnote{The meaning of this terminology
will become clear later.}
\begin{equation}
\Sigma(Y,\overline{Y},p,q) := {\cal F}(Y,\overline{Y})
- q_I (Y^I + \overline{Y}^I) + p^I ( F_I + \overline{F}_I) \;,
\label{EntropyFunction}
\end{equation}
where 
\begin{equation}
{\cal F}(Y, \overline{Y}) = -i (F_I \overline{Y}^I - Y^I \overline{F}_I)\;,
\end{equation}
is the black hole `free energy'. Both entropy function and free energy
are symplectic functions. It is straightforward to verify that
the extremisation equations 
\begin{equation}
\frac{\partial \Sigma}{\partial Y^I} = 0 = \frac{\partial \Sigma}{\partial \overline{Y}^I}
\end{equation} 
for the entropy function are precisely the black hole attractor
equations (\ref{AttractorEqsI}). Moreover, at the critical point $\Sigma$
equals the black hole entropy, up to a conventional factor:
\begin{equation}
\pi \Sigma_* = S_{\rm macro}(p,q) \;.
\end{equation}
The geometrical meaning of the variational principle becomes clearer
when one parametrizes the scalar manifold ${\cal N}_{\rm VM}$
using real variables \cite{CdWKM:2006}. 
Equation (\ref{EntropyFunction}) suggests
to use 
\begin{equation}
x^I = \mbox{Re} (Y^I + \overline{Y}^I) \;,\;\;\;
y_I = \mbox{Re} (F_I + \overline{F}_I)
\end{equation}
instead of $Y^I$ as coordinates on ${\cal N}_{\rm VM}$. 
The real quantities $x^I,y_I$ are indeed Darboux
coordinates on ${\cal N}_{\rm VM}$ \cite{Freed}.
The transition between the coordinates $Y^I = \mbox{Re} Y^I + 
i \mbox{Im} Y^I$ and the coordinates $x^I,y_I$ can be viewed
as a Legendre transform, which replaces $u^I= \mbox{Im} Y^I$ by
$y_I=\mbox{Re} F_I$ as independent variables. 
When working with
the real coordinates $x^I, y_I$, the metric on ${\cal N}_{\rm VM}$
(and in fact all couplings in the Lagrangian) can be expressed
in terms 
of a Hesse potential $H(x,y)$,\footnote{A Hesse potential
is the real analogue of a K\"ahler potential.} which is, up to 
a factor, the Legendre transform of the imaginary part
of the holomorphic prepotential \cite{Cor:01}:
\begin{equation}
H(x,y) = 2 \mbox{Im} F(x+iu) - 2 y_I u^I \;.
\end{equation}
Using the homogenity properties of the prepotential one 
sees immediately that the black hole free energy equals the
Hesse potential, up to a factor:
\begin{equation}
2 H (x,y) = {\cal F}(Y,\overline{Y}) = 
-i (F_I \overline{Y}^I - Y^I \overline{F}_I) \;.
\end{equation}
Moreover, when expressing the entropy function in terms of
real variables,
\begin{equation}
\Sigma(x,y,q,p) = 2 H (x,y) - 2 q_I x^I + 2 p^I y_I  \;,
\label{EntropyFctHesse}
\end{equation}
we see that the black hole variational principle also 
can be expressed in terms of a Legendre transform:
the black hole entropy is (up to a factor), the Legendre
transform of the Hesse potential
\begin{equation}
S_{\rm macro}(p,q) = 2 \pi \left(
H - x^I \frac{\partial H}{\partial x^I} - y_I \frac{ \partial H }{\partial y_I}
\right)_* \;.
\end{equation}
This observation motivates to call ${\cal F}(Y,\overline{Y})
\simeq H(x,y)$ the black hole free energy.

\subsection{$R^2$-corrections}

So far, our discussion referred to supergravity actions containing
up to two derivatives of the fields. Effective actions derived
from string theory contain an infinite series of higher derivative
terms, which are computable, at least in principle, in perturbation
theory. In four-dimensional $N=2$ supergravity there is a particular
class of higher derivative terms, often called `$R^2$-terms', 
for which an off-shell description is available. 
It is advisable to work in the superconformal formalism, where the
gravitational degrees of freedom reside in the Weyl multiplet. 
Then, $R^2$-terms can be taken into account by giving the 
prepotential an explicit dependence on an additional complex
variable $\Upsilon$, which is proportional to the lowest component
of the Weyl multiplet.\footnote{Since we work with the fields
$Y^I$ instead of $X^I$, we have applied an analogous rescaling.}
The function $F(Y^I, \Upsilon)$ is restricted by supersymmetry to
be holomorphic and (`graded') homogenous of degree two:
\begin{equation}
F( \lambda Y^I, \lambda^2 \Upsilon) = \lambda^2 F(Y^I, \Upsilon) \;. 
\end{equation}

In order to determine this function for $N=2$ string compactifications,
one expands it in powers of $\Upsilon$:
\begin{equation}
F(Y^I, \Upsilon) =\sum_{g=0}^\infty F^{(g)}(Y^I) \Upsilon^g  \;.
\end{equation}
It turns out that in type-II Calabi-Yau compactifications
the coefficient functions $F^{(g)}(Y^I)$ 
are proportional to the so-called genus-g topological 
free energy. I.e., $\exp F^{(g)}$ is the genus-g partition 
function of the topologically twisted type-II string on the
given Calabi-Yau manifold \cite{BCO}. In the string effective
action they occur as couplings in front of certain higher-derivative
terms \cite{AGNT}. Among these are terms of the form
\begin{equation}
{\cal L} \sim \sum_{g=1}^\infty \left( 
F^{(g)}(Y^I) (C^-_{\mu \nu \rho \sigma})^2 (T^-_{\alpha \beta})^{2g-2} 
+ \mbox{ c.c. } \right) + \cdots \;,
\label{HDterms}
\end{equation}
where $C^-_{\mu \nu \rho \sigma}$ is the anti-selfdual part of
the (space-time) Weyl tensor, and $T^-_{\alpha \beta}$ is
the anti-selfdual part of the antisymmetric tensor field sitting
in the lowest component of the Weyl multiplet.\footnote{By its
equation of motion, this field is related to graviphoton, and therefore
it is often simply referred to as the graviphoton.}
These terms are responsible for the terminology `$R^2$-terms',
as they are quadratic in the Riemann tensor. Note that
there are several other higher-derivative terms, which are
related to the ones given here by supersymmetry.

Using the superconformal off-shell formalism, the effect of these
$R^2$-terms on black holes can be analyzed in surprising generality
\cite{CdWM:98,CdWKM:2000}. One can even construct
explicit solutions, at least iteratively. 
The attractor equations generalize by replacing the prepotential
by the function $F(Y^I, \Upsilon)$, where the auxiliary field
$\Upsilon$ takes a specific constant value on the horizon:
\begin{equation}
\left( \begin{array}{c}
Y^I - \overline{Y}^I \\
F_I(Y,\Upsilon ) - \overline{F}_I(\overline{Y}, \overline{\Upsilon}) \\
\end{array}
\right)_* = i 
\left( \begin{array}{c}
p^I \\
q_I \\
\end{array}
\right) \;,\;\;\; \Upsilon_* = -64 \;.
\end{equation}
When computing the entropy one needs to replace the naive
area law by Wald's generalized formula (\ref{Wald}), since higher curvature
terms are present in the Lagrangian. Application of the general
entropy formula to the $N=2$ supergravity Lagrangian results
in the following expression for the entropy, which is a symplectic
function:
\begin{equation}
S_{\rm macro} = S_{\rm Wald} = 
\pi \left( (p^I F_I (Y,\Upsilon) - q_I Y^I ) - 
256 \; \mbox{Im} \left( \frac{\partial F}{\partial \Upsilon} \right) 
\right)_*  \;.
\end{equation}
Note that the entropy gets modified by the higher derivative terms
in two ways. First, the {\em area} of the event horizon changes, 
which is apparent from the presence of the full function $F(Y^I, \Upsilon)$
in the attractor equations and in the area term in the entropy 
formula. Second, there is a modification of the area {\em law} by the
last term. Like the area, this term is a symplectic function. 
A matching between macroscopic and microscopic entropy is only
found, when both corrections are taken into account \cite{CdWM:98}.

\subsection{Other higher derivative terms}

The Weyl multiplet encodes one specific class of higher derivative
terms, namely those of the form (\ref{HDterms}) and their
supersymmetry transforms. The string effective action includes
many other higher derivative terms, for example terms involving
higher powers of the Riemann tensor. It is natural to expect that
such terms should contribute to the entropy as well. 

There is, however, strong evidence that the $R^2$-terms are at least
the most important class. 
It is precisely these terms which are captured by the topological 
string. Since the topological string seems to encode 
all important phenomenological features of string compactifications,
one might expect that the same is true for black holes.
While this is not a particularly strong argument by itself,
strong `empirical' evidence
stems from the observation that already the inclusion of the
$R^2$-terms leads to a remarkable quantitative matching between
macroscopic and microscopic entropy, including 
in some cases even infinitely many subleading terms 
\cite{CdWM:98,CdWM:9906,CdWKM:2000}.
$R^2$-terms also have the remarkable property that they
can resolve null singularities \cite{DabKalMal}. These are singularities which
coincide with an event horizon. In our context this occurs 
in the form of black hole solutions which have a vanishing
area of event horizon. There are many examples where 
black hole solutions are null singular at the two-derivative
level but become regular black hole solutions with a finite
event horizon once $R^2$-corrections are taken into account.
Such black holes are called `small' black holes, because their
area is small when measured in string units. In contrast 
`large' black holes already have a nonvanishing horizon
at the two-derivative level. While $R^2$-terms resolve the
null singularities of most of the known small black holes, 
there are a few counterexamples
where further higher derivative terms are needed 
\cite{Sen:0504,DDMP2}. This illustrates that it 
is important to construct and study further classes of higher
derivative terms in supergravity. An important step forward
in this direction was made recently, in the framework of the
superconformal calculus \cite{deWSau}. One interesting 
observation made there is that while several 
further supersymmetric higher derivative terms can be constructed,
there are cancellations among them when the background is restricted
to be supersymmetric. This confirms the distinguished role 
of the $R^2$-terms. Another important observation
was made in \cite{BCM} and \cite{Sen:05}: the same answer
for the entropy is obtained by simply adding, instead of the whole
set of supersymmetric $R^2$-terms, the Gauss-Bonnet term to the
Einstein-Hilbert action. This suggest that the black hole entropy 
is a `robust' quantity, in the sense that gravitational Lagrangians
form large equivalence classes which lead to identical 
results for the entropy. Finally, yet another argument for the
particular role of the $R^2$-terms comes from the 
$AdS^3/CFT_2$ correspondence, on which we comment in the next
section.

\subsection{The $AdS^3/CFT_2$ correspondence}

The $AdS^3/CFT_2$ correspondence provides a general and
robust way to study the relation between microscopic and
macroscopic properties of four-dimensional black holes \cite{KraLar:2005,
GSY:06}. 
More generally, the correspondence can be 
applied whenever the near horizon
geometry of a black hole is locally of the form $AdS^3 \times S^n$,
where $S^n$ is an $n$-dimensional sphere. In the known examples
the global geometry of the three-dimensional part is a discrete
quotient of $AdS^3$, typically a BTZ black hole or an $SL(2,\mathbbm{Z})$
transform thereof \cite{KraLar:2006,dBCDMV}.  
For four-dimensional
extremal black holes the near horizon geometry is locally 
$AdS^2 \times S^2$, but in string compactifications the $AdS^2$
factor can combine with an internal $S^1$ to form an $AdS^3$.
One can then use the correspondence between gravity or string
theory on $AdS^3$ and a two-dimensional conformal field theory
on the (conformal) boundary of the $AdS$-space. States can
be counted in the $CFT_2$. Already the central charge
gives the leading asymptotics of states by Cardy's formula.
If the partition function is known, one can 
compute further corrections. Accepting the $AdS/CFT$ correspondence
in generality, one can apply it to non-superysmmetric
black holes. One can also
treat non-extremal black holes, if they can be related to 
non-extremal BTZ black holes.

In \cite{SaiSod,KraLar:2005} the correspondence was used to argue that
the Cardy formula for the $CFT_2$ agrees with the entropy of
the corresponding black hole, including higher derivative corrections.
Moreover, after lifting four-dimensional black holes to five dimensions
one can use anomalies to rederive
the entropy formula for BPS black holes in Calabi-Yau
compactifications of eleven-dimensional M-theory 
\cite{KraLar:2005,KraLar:0508}. This derivation
shows again the distinguished role of $R^2$-terms, since it suffices
to consider them in order to find the full anomaly. Thus other
higher derivative terms do not contribute to the anomaly, and, hence
not to the black hole entropy. This anomaly argument also applies
to non-BPS black holes.

The interested reader is referred to the nice lecture notes
\cite{Kraus} for a detailed review of the $AdS^3/CFT_2$ correspondence
and its application to black holes. If the partition 
function of the $CFT_2$ is known explicitly, one can evaluate the state
degeneracy beyond leading order. We will encounter some of these results
later in the context of black hole partition functions. Here we add
some comments on the limitations of the $AdS^3/CFT_2$ approach. 
The first limitation is that not all black holes can be related to
$AdS^3$. One straightforward way to obtain a four-dimensional extremal 
black hole with horizon $AdS^2 \times S^2$ from five dimensions 
is to compactify a five-dimensional BPS black string 
with horizon
$AdS^3 \times S^2$ \cite{ChaFerGibKal,ChaSab} 
on $S^1$, while adding momentum along the 
string \cite{MSW}. 
This accounts for half of the possible charges of a four-dimensional
black hole, and further charges can be included, for example
as induced M2-brane charges if one considers eleven-dimensional
supergravity compactified on $CY_3 \times S^1$ (where $CY_3$ denotes
a Calabi-Yau threefold) \cite{MSW}. In fact, 
even the general four-dimensional
BPS black hole can be lifted to a spinning five-dimensional BPS
black hole \cite{GSY:05}, but the near horizon geometry is more complicated.
In particular, one needs to superimpose a Taub-NUT
solution to account for non-vanishing D6 charge. Thus additional
input is needed to fully account for the microscopic description
of spinning five-dimensional and general four-dimensional BPS
black holes. Additional examples of interesting black holes which
do not have an $AdS^3$ description can be found in \cite{DabSenTri}.
Another problem is the regime in parameter space where one can
apply the correspondence after lifting a local $AdS^2$ factor to
a local $AdS^3$ factor. As explained in \cite{DabSenTri}
this is only adaequate if the additional circle is large, which
imposes a constraint on the charges. If, for example, all charges
are large, then the extra circle is not large, and the dimensional
reduction of the $AdS^3$ theory on the circle fails to capture
the full dynamics in $AdS^2$. As a consequence there are corrections
to the Cardy formula, which are sensitive to the details of the
boundary $CFT_2$. A probably related question is how to account
for the world-sheet and space-time instanton corrections of the
four-dimensional theory. This manifests itself already at the
two-derivative level. While supersymmetry only restricts 
the prepotential of the four-dimensional
supergravity theory to be holomorphic
and homogenous of degree two, 
the prepotential of the dimensionally
lifted five-dimensional supergravity theory must be a cubic
polynomial. The finiteness of the fifth dimension gives rise to
an infinite series of subleading corrections, which, in string 
theory, correspond to world-sheet and space-time instantons which
have finite action once the additional compact direction becomes
available. Finally, while the $AdS^3/CFT_2$ correspondence 
is robust in the sense of being independent of details of the 
dynamics, and can potentially be used for non-supersymmetric and
non-extremal black holes, one still needs to make the {\em assumption}
that the $AdS^2$ or $AdS^3$ geometry survives corrections, i.p. higher
derivative corrections. For four-dimensional BPS black holes
we know by explicit construction that the near horizon geometry
survives the $R^2$-corrections \cite{CdWKM:2000}, and it is
desirable to also have explicit results for five-dimensional black holes,
and for other classes of higher derivative terms.\footnote{Some of the 
results on five-dimensional black holes, which were originally
derived using the $AdS^3/CFT_2$ correspondence, have recently been
confirmed by direct computation (including higher derivative terms)
\cite{CDKL}. Upon dimensional reduction, these results agree with
those obtained direcly in four dimension.}

\subsection{Non-supersymmetric black holes}

The black hole 
attractor mechanism also works for black holes which are
extremal, but not supersymmetric, i.e. for black holes
which do not admit Killing spinors, but have
the near horizon asymptotics
$AdS^2 \times S^2$ \cite{FGK,GIJT}. With hindsight the presence of an attractor
mechanism is easy to understand, given the interpreation of
$AdS^2 \times S^2$ as a flux compactification. The background
flux induces an effective potential for the scalars, which 
is extremized at the horizon. This naturally fixes
the scalars.\footnote{If the potential 
has flat directions, then only part of the moduli get 
fixed.}  Since this argument relies on the field equations 
and the structure of the bosonic Lagrangian,
it does not depend on supersymmetry. It applies to 
non-supersymmetric Lagrangians, and to both BPS and non-BPS 
extremal solutions of supersymmetric Lagrangians.

An elegant formalism for the treatment of general extremal
black holes was introduced by A. Sen \cite{SenEntropyFct}. 
In this formalism one can use
any covariant Einstein-Maxwell type Lagrangian, including
matter and higher derivative terms, as input. From the
Lagrangian one extracts an entropy function by dimensional
reduction in the near-horizon geometry. The entropy function
yields the entropy upon extremisation. Since the entropy
can itself be expressed in terms of variational derivatives
of the Lagrangian, everything is tied directly to the
Lagrangian.  This approach is independent of the 
details of the field equations, as long as they come
from a generally covariant Lagrangian and admit 
extremal solutions. Since there
is an underlying variational principle, one might expect
that there is a close relation to the variational principle
for BPS black holes, which we described in section 
\ref{Sect:VarPri}. Indeed, it has been shown in \cite{CdWMa} 
that the two entropy
functions only differ by terms which vanish in supersymmetric
backgrounds.

The entropy function approach was used in \cite{SahSen}
to compute the entropy of BPS and non-BPS extremal black holes 
in $N=2$ supergravity with higher derivative corrections.
While the results agree with previous results for BPS black holes,
they found a disagreement for non-BPS black holes when comparing
to the dimensional reduction of the results of \cite{KraLar:2005}.
This discrepancy was resolved in \cite{SahSen:0608}, who found
that further terms in the effective action have to be taken
into account, which descend by dimensional reduction from 
Chern-Simons terms. 
While the original definition of an entropy function does not apply
when Chern-Simons terms are present, generalizations have been
formulated,
which allow to include three-dimensional BTZ black holes
\cite{SahSen:06} and  
extremal, rotating  black hole solutions
of five-dimensional supergravity \cite{COP,GolJen}.
An entropy function for non-extremal black holes with a
BTZ factor
has been proposed in \cite{SahSen:06,CaiPang}.

The detailed study of explicit non-BPS solutions
in supersymmetric compactifications, and of non-BPS fixed points
of the attractor equations has become an active subject
starting from  \cite{GIJT,Kal:2005}.
A very important question, which has been 
re-addressed recently in \cite{DabSenTri} is whether one can 
count microstates for non-BPS extremal black holes. 
Surprisingly the conclusion is that under certain
assumptions the attractor mechanism allows one to reliably
extrapolate the macroscopic entropy from the semi-classical gravity 
regime to the string perturbative regime, where microstates can be counted.

Finally, we already mentioned how one particular version of the
$AdS/CFT$ correspondence can be used to investigate non-BPS black
holes. Let us add that this correspondence can also applied 
in many other ways. In particular, it can be applied to 
non-extremal black holes, if they are asymptotically 
$AdS$ {\em at infinity}. Such black holes correspond to thermal states
of the dual gauge theory \cite{WitAdS}.

\section{Black hole partition functions}

\subsection{Reduced variational principles \label{Sec:RedVarPri}}

When we discussed the variational principle for BPS black holes 
in section \ref{Sect:VarPri}, all attractor equations 
were imposed simultanously. This procedure can be broken 
up into several steps, by imposing part of the attractor
equations and substituting them back into the entropy function.
If the resulting entropy function produces upon extremization
the remaining attractor equations, one has found a new, reduced
variational principle \cite{CdWKM:2006}. 
This works, in particular, if one imposes
the magnetic attractor equations in the first step, and this allows
us to relate the variational principle of section \ref{Sect:VarPri}
to the work of Ooguri, Strominger and Vafa
\cite{OSV} on black hole partition functions and the topological
string.

The magnetic attractor equations read  $Y^I - \overline{Y}^I =  i p^I$, 
and can be solved by setting
\begin{equation}
Y^I = \frac{1}{2} ( \phi^I + i p^I ) \;.
\label{solve_magnetic}
\end{equation}
The quantities $\phi^I = \frac{1}{2} \mbox{Re} Y^I =
\frac{1}{2} x^I$ are to be determined by the remaining, 
electric attractor equations. Looking at the gauge field
equations of motion one realizes that the $\phi^I$ are,
in the backgrounds under consideration, the electrostatic
potentials (see for example \cite{CdWKM:2000}). 
In thermodynamical terms this means that they are
the chemical potentials dual to the electric charge. 
Substituting the magnetic attractor equations into the entropy
function, one obtains the reduced entropy function
\begin{equation}
\Sigma (\phi, p,q ) = {\cal F}_{OSV}(p, \phi) - q_I \phi^I \;,
\end{equation}
where
\begin{equation}
{\cal F}_{OSV}(p,\phi) = 4 \mbox{Im} \left( F(Y,\Upsilon) \right)  \;,
\end{equation}
and $Y^I=Y^I(p,\phi)$ is given by (\ref{solve_magnetic}). 
Variation of the reduced entropy function yields 
the electric attractor equations
\begin{equation}
\frac{\partial {\cal F}_{OSV}}{\partial \phi^I} = q_I \;,
\end{equation}
and at the critical point one finds
\begin{equation}
S_{\rm macro}(p,q) = \pi \Sigma_* = \pi \left( {\cal F}_{OSV} -
\phi^I \frac{\partial {\cal F}_{OSV}}{\partial \phi^I} \right)_* \;,
\end{equation}
which shows that the black hole entropy is obtained from 
the `free energy' ${\cal F}_{\rm OSV}(p,\phi)$ through a
partial Legendre transform which replaces $\phi^I = \frac{1}{2}
\mbox{Re} Y^I$ by $q^I$ as independent variable.

The interpretation of ${\cal F}_{\rm OSV}(p,\phi)$ as a black hole
free energy is strongly supported by the following observation \cite{OSV}.
The function $F(Y^I, \Upsilon)$, which encodes the $R^2$-couplings 
together with the prepotential, is proportional to the (holomorphic)
free energy $F_{\rm top}$ of the topologically twisted type-II string.
Taking into account conventional normalization factors, the precise
relation is
\begin{equation}
e^{\pi {\cal F}_{\rm OSV}} = |e^{ F_{\rm top} } |^2 = | Z_{\rm top} |^2 \;,
\end{equation}
where $Z_{\rm top}$ is the (holomorphic) partition function of the
topological string. This suggests to take the interpretation
of ${\cal F}_{\rm OSV}(p,\phi)$ as a black hole free energy seriously.
Since this function depends on the magnetic charges $p^I$ and the
electrostatic potentials $\phi^I$, the free energy ${\cal F}_{\rm OSV}(p,
\phi)$ belongs to a 
mixed ensemble, where magnetic charges are treated microcanonically,
whereas electric charges are treated canonically. The partition function
for such an ensemble is
\begin{equation}
Z_{BH}(p,\phi) = \sum_{q} d(p,q) e^{\pi q_I \phi^I} \;,
\label{ZBH}
\end{equation}
where $d(p,q)$ is the microscopic state degeneracy, i.e., the
partition function of the microcanonical ensemble where both 
electric and magnetic charges are kept fixed. The two partition 
functions are related by a (discrete) Laplace transformation.

One can now formulated the OSV conjecture \cite{OSV} 
\begin{equation}
e^{\pi {\cal F}_{OSV}} \simeq Z_{BH}(p,\phi)  \;,
\label{OSVconjecture}
\end{equation}
where $\simeq$ denotes asymptotic equality in the limit of
large charges, which is the semi-classical limit. 
Thus the macroscopic quantity ${\cal F}_{\rm OSV}$, which is
determined by the couplings in the low energy effective action 
is directly related to the microscopic state degeneracy. 
Another, suggestive way to formulate the conjecture is \cite{OSV}
\begin{equation}
Z_{BH} \simeq |Z_{\rm top}|^2 \;.
\label{Z=Z2}
\end{equation}
By only requiring asymptotic equality we have formulated a
`weak' version of the OSV conjecture \cite{OSV}. Here
asymptotic equality means equality order by order in an 
expansion in inverse charges, where the charges are taken to
be uniformly large. Due to the homogenity properties of the
prepotential, such an expansion is natural. The weak form of 
the conjecture has been tested successfully for `large' black 
holes, while there are problems with the subleading terms for
`small' black holes. We will come back to this later.

An even more intriguing `strong' version of the conjecture
\cite{OSV}
asserts that (\ref{Z=Z2}) can be extended to an 
exact equality, which might
even provide a non-perturbative definition of the topological
string in terms of black hole data \cite{OSV,DGNV}.
As we will discuss below,
the status of this version is much less clear. In any
case the original conjecture needs to be amended. In particular,
the holomorphic factorization (\ref{Z=Z2}) cannot be exact.

Tests of the conjecture can be performed by either predicting 
the free energy ${\cal F}_{\rm OSV}(p,\phi)$ from the microscopic
state degeneracy $d(p,q)$, or vice versa. For the latter method
one formally inverts (\ref{ZBH}) and expresses the state degeneracy
through an inverse Laplace transformation:
\begin{equation}
d(p,q) \simeq  \int_C  d \phi^I  \; e^{ \pi {\cal F}_{OSV} - q_I \phi^I} \;.
\label{InverseLaplace}
\end{equation}

At this point we should mention several open questions.
One point is that the sums and integrals
appearing in (\ref{ZBH}), (\ref{InverseLaplace}) are given
as formal expressions. The leading term in the integral
(\ref{InverseLaplace}) is a Gaussian integral associated to
an indefinite quadratic form, which requires resummation or
analytical continuation. The convergence of 
(\ref{ZBH}), (\ref{InverseLaplace}) and the proper choice 
of integration contours in (\ref{InverseLaplace}) have not been 
investigated in much detail. Also note that $\phi^I$ appears
as a complex variable in (\ref{InverseLaplace}), while its
physical values are real, which shows again that the definition
of (\ref{InverseLaplace}) involves an analytic continuation.
These questions can be ignored as long as
(\ref{ZBH}), (\ref{InverseLaplace}) are evaluated in saddle point
approximation, but become relevant if one wants to go beyond this.

Another noteworthy point is that symplectic covariance 
(electric-magnetic duality) is not manifest in the OSV conjecture. Since 
stringy symmetries, such as S-duality and T-duality form a 
discrete subgroup of the symplectic rotations, one might be 
worried whether the conjecture is compatible with string dualities.
A closely related point is the role of the so-called   
non-holomorphic corrections, to be discussed below. There we will
formulate a manifestly duality invariant version of the OSV
conjecture, which predicts the presence of a specific
correction factor in (\ref{ZBH}), (\ref{InverseLaplace}).

There is also the question whether $d(p,q)$ really is
the absolute state degeneracy, or a weighted, index-like quantity.
Here we should remember that we need to vary parameters to 
go from the string perturbative regime (state counting) to the
effective gravity regime (black hole with event horizon). 
During the interpolation, BPS states might pair up into non-BPS
states, or decay when crossing lines of marginal stability. 
Index-like quantities, like the elliptic genus, are at least
insensitive to the first effect. Therefore it appears natural from
the microscopic point of view that $d(p,q)$ is an index
\cite{OSV}. However,
entropy is normally related with the absolute state degeneracy.
An extensive study of `small' black holes in various $N=4$ and
$N=2$ compactifications \cite{DDMP1,DDMP2}
found that most cases are insensitive
to the distinction between an indexed and an absolute degeneracy.
However, for untwisted states in $N=2$ orbifolds the OSV
prediction for the state degeneracy agrees with the absolute
degeneracy and disagrees with the exponentially smaller indexed
degeneracy \cite{DDMP2}. Related observations were reported
and discussed in \cite{Sen:0504}. Recently \cite{DenMoo}
have made an interesting proposal for resolving the problem:
they suggest that the true degeneracy, when computed at
{\em finite} coupling, is always counted by an appropriate index. 
This idea makes use of the fact that BPS states can become
marginally stable and decay upon variation of the coupling.


\subsection{Non-holomorphic corrections}

So far, our discussion of $N=2$ BPS black holes was based
on the holomorphic function $F(Y^I, \Upsilon)$. This description
is however incomplete. This is seen immediately when looking
at explicit examples of black hole solutions. A particularly
instructive class of examples is provided by compactifications
with enhanced $N=4$ supersymmetry. These can be described in 
terms of the $N=2$ formalism used here. The function 
$F(Y^I, \Upsilon)$ takes the special form
\begin{equation}
F(Y^I, \Upsilon) =  - \frac{ Y^1 Y^a \eta_{ab} Y^b}{Y^0} 
+ F^{(1)}(S) \Upsilon \;.
\end{equation}
Here $a=2, \ldots, n$, where $n$ is the number of 
$N=2$ vector multiplets. In heterotic $N=4$ compactifications
$S = -i \frac{Y^1}{Y^0}$ is the dilaton. 
Note that, in contrast to the $R^2$-coupling $F^{(1)}$, 
the prepotential does not receive 
instanton corrections, and that all the higher coupling 
functions $F^{(g)}(Y^I)$, $g>1$ vanish. Moreover S- and T-duality
are believed to be exact symmetries in $N=4$ compactifications.
The heterotic dilaton $S$ is inert under T-duality and transforms
fractionally linear under S-duality
\begin{equation}
S \rightarrow \frac{ aS -ib}{icS +d} \;.
\end{equation}
It is then straightforward to show that the attractor equations
and the black hole entropy can only be S-duality covariant if
the function $f(S) = - i \frac{\partial F^{(1)}(S)}{\partial S}$
transforms with weight 2 \cite{CdWM:9906}. 
But this is not possible if $F^{(1)}(S)$
is holomorphic. The only way to get an S-duality invariant entropy
is to add non-holomorphic terms. Both the problem and its solution
are variants of a well known phenomenon, which occurs generally 
in supersymmetric string effective field theories \cite{DKL}. One has
to distinguish between two types of couplings: the Wilsonian
couplings are part of a local Wilsonian effective action. In 
supersymmetric theories this results in a holomorphic dependence
on the moduli. However the Wilsonian couplings are different
from the physical couplings which can be extracted from string
scattering amplitudes. In particular, they do not transform 
covariantly under string dualities. In contrast, the physical
couplings are duality covariant, but have a more complicated,
non-holomorphic dependence on the moduli. They are related 
to a different type of effective action, the generating functional
of one particle-irreducible graphs. If the theory contains
massless particles, this type of action is in 
general non-local.

Since the black hole entropy is a physical quantity, it needs
the full physical couplings as input. A systematic way of 
incorporating non-holomorphic corrections into the attractor
equations, the variational principle and the black hole entropy 
was worked out in \cite{CdWM:9906,CdWKM:2004,CdWKM:2006}.
Basically, the modification amounts to the following replacement:
\begin{equation}
\mbox{Im}(F(Y,\Upsilon)) \rightarrow \mbox{Im} (F(Y,\Upsilon))
+ 2 \Omega(Y,\overline{Y}, \Upsilon, \overline{\Upsilon}) \;,
\end{equation}
where $\Omega(Y,\overline{Y}, \Upsilon, \overline{\Upsilon})$ is 
a real-valued function which is homogenous of degree two. For 
concrete models this function has to be computed in string theory.
Note that any harmonic part of $\Omega$ can be absorbed into 
$\mbox{Im}F$. In other words, `non-holomorphic' corrections 
correspond to non-harmonic functions $\Omega$. 

The non-holomorphic contributions to the string effective 
action have their microscopic counterpart in the holomorphic
anomaly of the topological type-II string \cite{BCO}. Consequences
of the holomorphic anomaly for the OSV conjecture have been
discussed in \cite{OSV,DGNV,Ver:2004,SenNH}. Our `macroscopic' 
approach, whose relation to 
these `microscopic' approaches is not completely understood, takes the 
non-holomorphic corrections into account ab intio, and 
maintains manifest symplectic covariance
and duality invariance throughout. This leads us to postulate
a specific modified version of (\ref{OSVconjecture}), which is based on 
the black hole variational principle of section \ref{Sect:VarPri}.
Consequently, we do not start from the mixed, but from the
canonical ensemble and postulate that the black hole
partition function is given by
\begin{equation}
e^{ 2 \pi H(x,y,\Upsilon, \overline{\Upsilon}) } 
\approx  Z_{\rm BH}^{\rm (can)} =
\sum_{p,q} d(p,q) e^{2 \pi [q_I x^I - p^I y_I ]}  \;,
\label{OSVmodHesse}
\end{equation}
or, equivalently, 
\begin{equation}
e^{\pi {\cal F}(Y,\overline{Y}, \Upsilon, \overline{\Upsilon})}  
\approx  Z_{\rm BH}^{\rm (can)} =
\sum_{p,q} d(p,q) e^{\pi [ q_I (Y^I + \overline{Y}^I)  - p^I
(F_I + \overline{F}_I) ]} \;.
\label{OSVmodFree}
\end{equation}
Here the generalized Hesse potential $H$ and the canonical free
energy ${\cal F}$ 
include the non-harmonic function $\Omega$. By an inverse 
Laplace transformation we can solve for the microstate degeneracy:
\begin{equation}
d(p,q) \approx \int dx dy e^{\pi \Sigma(x,y)} 
\approx \int dY d \overline{Y} \Delta^-(Y, \overline{Y}) 
e^{\pi \Sigma (Y, \overline{Y})} \;, 
\label{StatesFromCan}
\end{equation}
where $\Sigma$ is the entropy function and
\begin{equation}
\Delta^{\pm}(Y,\overline{Y}) = 
| \det \left[
\mbox{Im} F_{KL} + 2 \mbox{Re}(\Omega_{KL} \pm \Omega_{K\overline{L}}) 
\right] | \;.
\end{equation}
When expressing the integral in terms of the real variables $x^I, y_I$, 
the measure is just the natural symplectically invariant measure
on ${\cal N}_{\rm VM}$ (proportional to the top exterior power of
the symplectic form). Conversion to the complex variables $Y^I$
results in a rather complicated measure factor. This shows that
the real variables are conceptually more natural. However, in practice 
we only know the subleading corrections to $\Sigma$ in terms of
complex variables. Fortunately, the measure factor is subleading
in the limit of large charges. The microscopic entropy is 
by definition the natural logarithm of the state degeneracy,
while the macrocopic entropy is given by the saddle point
value of the entropy function, $S_{\rm macro}(p,q) = \pi \Sigma_*$. 
If we evaluate (\ref{StatesFromCan}) in saddle point approximation,
we obtain the following relation:
\begin{equation}
d(p,q) = e^{S_{\rm micro}} 
\approx e^{S_{\rm macro}(p,q)} \sqrt{ \frac{\Delta^-}{\Delta^+}}
\approx e^{S_{\rm macro}(p,q) (1 + \cdots)} \;.
\end{equation}
Here $\Delta^+$ is the fluctuation determinant of the Gaussian
integral around the stationary point. Both entropies are different
in general, as expected, since they correspond to 
different ensembles. Since the correction factors $\Delta^{\pm}$
are subleading, they agree in the limit of large charges,
which plays the role of the thermodynamic limit. 

It is instructive to evaluate (\ref{StatesFromCan}) in such a
way that one can compare to the original OSV conjecture (\ref{OSVconjecture}).
The observation that the variational principle can be broken
up into two steps can be used to perform a saddle
point evaluation of (\ref{StatesFromCan}) with respect to
$\mbox{Im}Y^I$, which gives
\begin{equation}
d(p,q) \approx \int d \phi 
\sqrt{\Delta^-(p,\phi)}
e^{\pi [ {\cal F}_E(\phi,p) - q_I \phi^I ] }  \,.
\label{StatesFromMod}
\end{equation}
Here we denote the free energy in the mixed ensemble by ${\cal F}_E$
instead of ${\cal F}_{\rm OSV}$ because we include the non-holomorphic
corrections. After applying a discrete Laplace transform
\begin{equation}
\sqrt{\Delta^-} e^{\pi {\cal F}_E(p,\phi)}  \approx 
Z^{\rm (mix)}_{\rm BH} = \sum_q d(p,q) e^{\pi q_I \phi^I}  \;,
\label{OSVmod}
\end{equation}
we can compare to the original OSV conjecture
\begin{equation}
e^{\pi {\cal F}_{OSV}(p,\phi)}  \approx 
Z^{\rm (mix)}_{\rm BH} = \sum_q d(p,q) e^{\pi q_I \phi^I} \;.
\end{equation}
We see that imposing symplectic covariance has lead us to two
modifications: we have an additional measure factor $\Delta^-$,
and we have included the non-holomorphic corrections.

\subsection{Large black holes}

We now turn to tests of our conjectures (\ref{OSVmodHesse})--(\ref{OSVmod})
about black hole partition functions. Here we concentrate on 
$N=4$ compactifications,
where exact S- and T-duality invariance and the vanishing of the
higher coupling functions $F^{(g)}$, $g>1$ leads to simplifications. 
In $N=4$ compactifications there are two types of BPS states:
$\frac{1}{4}$-BPS states which carry both electric and magnetic 
charge and $\frac{1}{2}$ BPS states which are purely electric
or purely magnetic.\footnote{This is meant modulo symplectic
rotations.} These correspond to large and to small black holes,
respectively.
In this section we consider $\frac{1}{4}$-BPS states,
for which the following formula for the state degeneracy
has been conjectured \cite{DVV,SSY}:
\begin{equation}
d(p,q) = \oint d\rho d \sigma d v
\frac{ e ^{i \pi  [ \rho p^2 + \sigma q^2 + (2v-1) pq)]}}
{ \Phi_{10}(\rho, \sigma, v) }  \;.
\end{equation}
The integral is a three-fold contour integral in the so-called
rank 2 Siegel upper half space, which can be representated as
the space of complex, symmetric matrices with positive imaginary
part,
\begin{equation}
\Omega = \left( \begin{array}{cc}
\rho & v \\ v & \sigma \\
\end{array} 
\right) \;,\;\;\; \rho_2  > 0 \;,\;\;
\sigma_2 > 0 \;,\;\;
\rho_2 \sigma_2 - v_2^2 > 0 \;.
\label{DVV}
\end{equation}
The function $\Phi_{10}(\rho, \sigma,v)$ is the weight 10 
Siegel cusp form, which generalizes the well-known 
discriminant function $\eta^{24}(\rho)$. $p$ and $q$ are the
vectors of electric and magnetic charges, and $p^2, q^2, 
p\cdot q$ are the three T-duality invariant scalars one
can form out of them. Since these three T-duality invariants
transform in a three-dimensional representation of the
S-duality group $SL(2,\mathbbm{Z})$, the above expression
for $d(p,q)$ is formally $S$- and $T$-duality invariant, as it must be.
Strong support for (\ref{DVV}) has been given by various independent
arguments, including derivations from the worldvolume theory
of type-II NS5-brane \cite{DVV} and from the D1-D5 bound state
\cite{SSY}. The formula also has been generalized to the so-called
CHL models \cite{CHL}, which are $N=4$ orbifolds, in \cite{JatSen}. 
The microscopic
formula (\ref{DVV}) can be evaluated in a saddle point approximation.
The result can then be compared to the macroscopic black hole
entropy and to the black hole partition function, including
subleading corrections \cite{CdWKM:2004}.

The key for handling the subleading corrections on the 
macroscopic side is the observation that for $N=4$ compactifications
there is a further reduction of the black hole variational
principle \cite{CdWM:98,CdWKM:2006}. Due to the 
special structure of the prepotential,
one can solve all but two attractor equations explicitly,
for general values of the charges. The remaining two attractor
equations then determine the value of the (heterotic) dilaton on
the horizon. Substituting the solution back into the entropy
function, one obtains the reduced entropy function
\begin{equation}
\Sigma(S, \overline{S},p,q) =
- \frac{q^2 - i p\cdot q (S - \overline{S}) + p^2 |S|^2}{S+\overline{S}}
+ 4 \Omega(S, \overline{S},\Upsilon, \overline{\Upsilon}) \;.
\label{EntropyFunctionDilatonic}
\end{equation}
Here we absorbed the holomorphic $R^2$-couplings $F^{(1)}$ into
the function $\Omega $
$=\Omega(S,\overline{S}, \Upsilon, \overline{\Upsilon})$
for convenience. 
Given the transformation properties of the charges and of the dilaton,
$\Sigma$ is a symplectic function, and the black hole entropy 
$S_{\rm macro}(p,q) = \pi \Sigma_*$ is manifestly S- and T-duality
invariant, for any S-duality invariant function $\Omega$.

At the two-derivative level, which corresponds to $\Omega =0$,
the BPS black hole entropy is
\begin{equation}
S_{\rm macro} = \pi \sqrt{q^2 p^2 - (q \cdot p)^2 }
\end{equation}
\cite{CveYou,CveTse,BerKalOrt}, which agrees with the saddle point value of 
(\ref{DVV}) \cite{DVV}. For the simplest $N=4$ compactification,
which can be realized by compactifying the heterotic string on 
$T^6$, $\Omega$ takes the form
\begin{equation}
\Omega = \frac{1}{256 \pi} \left[
\Upsilon \log \eta^{24}(S) + 
\overline{\Upsilon} \log \eta^{24}(\overline{S}) +
\frac{1}{2}  ( \Upsilon + \overline{\Upsilon} ) \log (S + \overline{S})^{12}
\right]
\label{Omega4R2NH}
\end{equation}
\cite{HarMor:96}. Expansion of the Dedekind function
\begin{equation}
\log \eta^{24}(S) = - 2 \pi S - 
24 e^{-2\pi S} - 36 e^{-4 \pi S} - 32 e^{-6 \pi S}
+ {\cal O}
(e^{-8 \pi S})
\end{equation}
shows that there is a tree-level $R^2$-term, which receives an 
infinite series of instanton corrections. The integral
(\ref{DVV}) is evaluated in two steps. The $v$-integral is done
by taking residues. It can be shown that for large dyonic charges
one residue dominates, while all others are exponentially 
surpressed \cite{DVV,ShiYin}. 
After taking this leading residue, the $\rho$- and $\sigma$-integrals
can be performed in saddle point approximation. It turns out that 
the saddle point equations for $\rho$ and $\sigma$ are precisely the
two attractor equations for the dilaton, which remain after the
other attractor equations have been solved \cite{CdWKM:2004}. 
As a result macroscopic
and microscopic entropy agree (the $\Delta^{\pm}$ factors cancel, possibly 
up to subleading terms). This result is semi-classical, as it is obtained
in saddle point approximation. However, the matching involves the full
function $\Omega$, including the infinite series of instanton corrections,
which are exponentially surpressed for large charges \cite{CdWKM:2004}.  

In order to draw conclusions about the OSV conjecture, one can
proceed in a different way, following \cite{ShiYin}. Namely, one
can use the integral representation (\ref{DVV}) for the
state degeneracy to evaluate the mixed partition function
$Z(p,\phi) = \sum_q d(p,q) e^{\pi q_I \phi^I}$. The result can be written
in the following way
\begin{equation}
Z(p,\phi) = \sum_{\rm shifts} \sqrt{ \tilde{\Delta}(p,\phi)}
e^{\pi {\cal F}_E(p,\phi)} \,.
\end{equation} 
The right hand side of this equation contains a sum over shifts 
in $\phi$. This has to be included to match the manifest periodicity
properties of the left hand side.\footnote{A systematic understanding
of periodicity properties is another aspect of the OSV conjecture which
deserves further attention. It is of course closely related to the questions
of analytic continuations and choices of integration contours, etc.,
which we mentioned at the end of  section \ref{Sec:RedVarPri}.}
The function in the exponent agrees exactly with the mixed ensemble 
black hole free energy
\begin{equation}
{\cal F}_E(p,\phi) = \frac{1}{2} ( S + \overline{S}) \left(
p^a \eta_{ab} p^b - \phi^a \eta_{ab} \phi^b \right)
-i (S - \overline{S}) p^a \eta_{ab} \phi^b + 4 \Omega(S, \overline{S}, 
\Upsilon, \overline{\Upsilon}) \;,
\end{equation}
including all $R^2$- and nonholomorphic terms \cite{CdWKM:2006}. 
Finally there is
a non-trivial measure factor $\tilde{\Delta}^-$, 
which agrees with our prediction  (\ref{OSVmod}) in the limit of large charges.

We can also connect our results to an observation made
in \cite{DavSen}. There it was shown that the microscopic 
state degeneracy for $N=4$ compactifications can be brought 
to the form
\begin{equation}
d(p,q) = \int \frac{d S d \overline{S}}{(S+\overline{S})^2}
\, e^{\pi \tilde{\Sigma} (S, \overline{S}, p,q)}  \;.
\label{MicroEntro}
\end{equation}
Both the function $\tilde{\Sigma}$ and the measure are
manifestly S- and T-duality invariant. The function $\tilde{\Sigma}$
can be interpreted as a microscopic entropy function, i.e., as
a microscopic  analogue of the macroscopic
entropy function used in the formalism 
of \cite{SenEntropyFct}. The state degeneracy can be evaluated systematically
by treating $\tilde{\Sigma}$ as a classical Lagrangian, adding
a source term, and using Feynman diagram and background field
techniques to evaluate the corresponding partition 
function \cite{DavSen}. 
While the tree level evaluation corresponds to taking the
saddle point value of the integral, the one loop level corresponds
to the full saddle point evaluation, including the fluctuation
determinant. We expect that at least in this approximation the
result should agree with the one obtained from our macroscopic
entropy function (\ref{EntropyFunction}) (generalized to 
include $R^2$- and non-holomorphic corrections \cite{CdWKM:2006}). 
To show that this is indeed the case, we note
that in the saddle point approximation $\tilde{\Sigma}$ takes
the following form:
\begin{eqnarray}
\tilde{\Sigma} &=& - \frac{ q^2 + i (S -\overline{S}) pq +
|S|^2 p^2 }{S + \overline{S}} - \frac{2}{\pi} \ln [ (S+\overline{S})^6 
|\eta(S)|^{24} ] \nonumber \\
&& + \frac{1}{\pi} \ln \left[ - \frac{ q^2 - i (S -\overline{S}) pq +
|S|^2 p^2 }{S + \overline{S}} \right] + \mbox{const.}  + {\cal O}{(1/Q)} \;,
\end{eqnarray}
which is manifestly S- and T-duality invariant.\footnote{${\cal O}{(1/Q)}$
indicates terms which involve inverse powers of the charges.} 
Here we converted a result of \cite{DavSen} into our conventions. Constant
terms and an infinite series of terms suppressed by negative powers of
the charges have been neglected. 
Looking back to (\ref{EntropyFunctionDilatonic}) 
and (\ref{Omega4R2NH}),
and taking into account that $\Upsilon_* = -64$ we immediately see
that 
\[
\tilde{\Sigma} = \Sigma(S,\overline{S},p,q) +\frac{1}{\pi} \ln \hat{K}_* +
\cdots \;,
\]
where $\Sigma$ is the reduced `dilatonic' entropy function 
(\ref{EntropyFunctionDilatonic}).
$\hat{K}_*$ is the attractor value of the symplectic function
$\hat{K} =
i(\overline{Y}^I \hat{F}_I - Y^I \overline{\hat{F}}_I)$, where
the `hat' signals the inclusion of non-holomorphic terms. At
the attractor point this quantity is
proportional to the area \cite{CdWKM:2000},\footnote{Here
$|Z|^2 = p^I F_I(Y\Upsilon) - q_I Y^I$ involves
the full function $F(Y,\Upsilon)$.}
\begin{equation}
\hat{K}_* = |Z|^2_* = \frac{A}{4\pi}= -
\frac{ q^2 + i (S -\overline{S}) pq +
|S|^2 p^2 }{S + \overline{S}} \;.
\end{equation}
The quantity $\hat{K}$ is the leading term of the measure $\Delta^-$
in the canonical partition function. 
In a saddle point approximation
one can bring the partition function (\ref{OSVmodFree})
to the following form, where
all but two integrations have been performed:\footnote{This formula is a 
consequence of the results of \cite{CdWKM:2006}.}.
\begin{equation}
d(p,q) \simeq \int \frac{dS d \overline{S}}{(S + \overline{S})^2}
\hat{K}_* e^{\pi \Sigma(S,\overline{S},p,q)}\;,
\end{equation}
which agrees with the microscopic formula (\ref{MicroEntro}).
Thus in a saddle point approximation 
the microscopic entropy function of \cite{DavSen} 
can be matched macroscopically  by absorbing the measure factor into 
the reduced dilatonic entropy 
function, which includes $R^2$ and non-holomorphic corrections.

The tests described in this section provide strong evidence that
our  proposed modifications of
the OSV conjecture are correct in the semi-classical limit.
Note that while we displayed explicit
results specifically for toroidal heterotic compactifications, the same 
tests are passed successfully by  CHL models \cite{CdWKM:2006}.
This generalisation relies on the explicit results on CHL models
obtained in \cite{JatSen}. 
The presence of a measure factor in the OSV formula has been
established beyond doubt \cite{ShiYin,CdWKM:2006}.\footnote{Further
indications that a measure factor is unavoidable come from
the study of small black holes, to be reviewed below.}
Note, however, that the explicit measure factor that we propose has
so far only been tested in the saddle point approximation.
In view of the intricate nature of further corrections it is possible,
in fact even quite likely that this is not the full story.

\subsection{Small black holes}

Due to lack of space we only discuss the situation
concerning small black holes very briefly. A more
detailed discussion can be found in \cite{DDMP1,DDMP2}
and in \cite{CdWKM:2006}. While many more examples
have been studied in \cite{DDMP1,DDMP2}, we consider
the simplest case, electric black holes in toroidal
compactifications of the heterotic string, which serves
us well to explain what we consider to be the main
problem. These black holes are expected to be in
one-to-one correspondence with the $\frac{1}{2}$-BPS
states of the perturbative heterotic string, known
as Dabholkar-Harvey states \cite{DabHar}. The problem
of computing the state degeneracy is formally equivalent
to (a version of) the classical problem of counting partitions
of integers, which was studied by Hardy and Ramanujan,
and which can be solved using a technique developed
by Rademacher \cite{Rademacher}.\footnote{See \cite{FareyTail1}
for a readable account.} Therefore this is 
an ideal case for precision tests of the OSV conjecture
\cite{Dabholkar}. Using the Rademacher expansion, 
the microscopic entropy is (see for example \cite{FareyTail1}):
\[
S_{\rm micro} \simeq \hat{I}_{13}\left(4 \pi \sqrt{ \frac{1}{2} |q^2| }
\right) \simeq 4 \pi \sqrt{ \frac{1}{2} |q^2|} - 
\frac{27}{4} \log |q^2| + \cdots 
\]
Here $q^2$ is the T-duality invariant norm of the vector 
$q=(q_0,p^1,q_2,\ldots)$ of electric charges\footnote{
$p^1$ is a magnetic charge in the `supergravity frame' used 
in this article, but an electric charge, carried by fundamental
heterotic strings, in the `string frame'.}, and
$\hat{I}_{13}$ is a modified Bessel function. The Rademacher
expansion contains an infinite series of further Bessel 
functions, which are exponentially surpressed for large charges.
We also have made explicit the first two terms in the
asymptotic expansion of the leading Bessel function. 
There is an infinite series of 
further terms involving inverse powers of the charges, see for example 
\cite{DDMP2}. In \cite{DDMP2} the prediction of the 
unmodified OSV conjecture for the state degeneracy was found 
to be
\begin{equation}
S_{\rm OSV} = 
(p^1)^2 \hat{I}_{13}\left(4 \pi \sqrt{ \frac{1}{2} |q^2| }
\right) \;.
\end{equation}
Here `unmodified OSV conjecture' means that the integral 
(\ref{InverseLaplace}) was evaluated without a measure
factor and without including non-holomorphic terms. 
This result matches the leading terms and almost matches
infinitely many subleading terms. However, the prefactor
$(p^1)^2$ spoils T-duality and leads to a mismatch
of the subleading terms. Moreover, in order to get a 
Bessel function with the correct index $13$, one needs
to truncate the OSV integral from 28 to 24 integrations
\cite{DDMP2}.\footnote{The 
dyonic case works with the full number
28 of integrations (i.e. one integration for each charge) \cite{CdWKM:2006}.}
The failure of T-duality suggests the presence of a 
measure factor in the OSV integral \cite{DDMP2}. When evaluating 
the modified OSV integral (\ref{StatesFromMod}),
which includes a measure factor and the non-holomorphic
terms, one finds \cite{CdWKM:2006}:
\begin{equation}
S_{\rm OSV, mod}
\simeq \hat{I}_{13-\frac{1}{2}}
\left(4 \pi \sqrt{ \frac{1}{2} |q^2| }
\right) \simeq 4 \pi \sqrt{ \frac{1}{2} |q^2|} - 
\frac{13}{2} \log |q^2| + \cdots 
\end{equation}
This is manifestly T-duality invariant, and the OSV integral
involves the full set of 28 integrations, as in the 
dyonic case. However, a contribution from the measure 
shifts the index of the Bessel function, which results in
a systematic mismatch of infinitely many subleading 
terms encoded in the leading Bessel function. This is
puzzling. Moreover, the structure of the exponentially
surpressed terms seen in any version of the OSV conjecture
is completely different from the one appearing in 
microscopic state counting \cite{DDMP2,CdWKM:2006}.

As a result the status of the OSV conjecture for electric
black holes is unclear. Only the leading term matches
quantitatively. One obvious reason for the problem is that
in the limit of large charges the (would-be) leading 
contribution to the entropy and to the measure
factor vanishes. Thus the starting point of the expanison
is ill defined. One might still hope to find an improved
expansion which leads to a matching for the inverse power
corrections. Exponentially surpressed corrections, if they
can be made work at all, seem
to require a drastic modification.

\subsection{Why $Z_{\rm BH} \not= |Z_{\rm top}|^2$.}

While it was clear from the beginning that the relation
$Z_{\rm BH} = |Z_{\rm top}|^2$, with $Z_{\rm top}$ 
being holomorphic, cannot be literally 
true due to existence of non-holomorphic corrections \cite{OSV},
this is not always appreciated. From the macroscopic
perspective the non-holomorphic contributions reflect that
the black hole partition function is a physical quantity,
because it encodes state degeneracies. Therefore it must
be related to the `physical' rather than the `Wilsonian'
couplings. In fact, the structure to be expected is 
familiar from supersymmetric mass formulae of the 
form $M^2 = e^K |W|^2$, which involve the square
of a holomorphic object (the superpotential $W$, evaluated
at its minimum) together with a non-holomorphic
factor, the exponentiated K\"ahler potential. Therefore the
presence of a non-holomorphic measure factor is not surprising. What is 
less clear is how this factor can be understood microscopically.

Recently, there has been progress in understanding
the subleading microscopic contributions to the entropy
of $N=2$ black holes, by combining various ingredients such as the
the $AdS^3/CFT_2$ correspondence, the elliptic genus, and 
generalizations of the Rademacher expansion. In particular, 
a microscopic explanation has been found why $Z_{\rm BH} \simeq
|Z_{\rm top}|^2$ in the limit of large charges
\cite{GSY:06,KraLar:2006,BGGHSY,dBCDMV}:
the asymptotic
factorization reflects that the entropy receives contributions
from both branes and anti-branes, which do not couple
to leading order. Therefore the
breakdown of holomorphic factorization should result from 
a breakdown of decoupling. Most recently, the refined 
analysis performed in \cite{DenMoo}
has indeed identified a microscopic measure factor,
which agrees, in the semiclassical limit, with the factor
found in \cite{ShiYin,CdWKM:2006}.

\subsection{Concluding remarks}

Supersymmetric black holes and their role in string theory
have been a topic of tremendous interest and activity over
the last years. This overview could not do justice to all
aspects of this field, and it has been biased in stressing
the macroscopic or supergravity perspective. A good introduction
to the microscopic aspects is included in the nice set of
lecture notes \cite{Pioline}, while \cite{Kraus} reviews
the $AdS^3/CFT_2$ perspective. 
We have tried to be 
complete at least in our discussion of open questions and problems,
since we believe that this will continue to be an exciting
topic of research for years to come.

\subsection*{Acknowledgement}
I would like to thank the organisers, and in particular
 Franco Pezzella, for organising a wonderful meeting and
for giving me the opportunity to talk at length about a
fascinating topic. My own contributions to the subject were
made in collaboration with Gabriel Lopes Cardoso, Bernard
de Wit and J\"urg K\"appeli. Special thanks go to Gabriel Lopes
Cardoso for various helpful remarks on the manuscript.

\end{document}